\documentclass[technote,draftcls,onecolumn,11pt,twoside]{IEEEtran}
\ifCLASSINFOpdf
\else
\fi
\usepackage[utf8]{inputenc}
\usepackage{cite}

\usepackage{latexsym,epsfig,amssymb,amsfonts,eucal,graphicx}
\usepackage{subfigure}
\usepackage[cmex10]{amsmath}
\usepackage{slantsc}
\usepackage{enumerate}
\usepackage{setspace}
\usepackage{graphicx}
\usepackage{epstopdf}
\usepackage{mdwmath}
\usepackage{mdwtab}
\usepackage{color}
\usepackage{soul}
\usepackage{multirow}
\usepackage{array}

\begin{document}
%
\title{Near-Field Perturbation Effect on Constellation Error in Beam-Space MIMO }
%
%
%

\author{Mohsen~Yousefbeiki,~
        Juan~R.~Mosig,~
        and~Andreas~Burg
\thanks{This work was supported by the Swiss National Science Foundation (SNSF) under Grant No. 133583.}
\thanks{M. Yousefbeiki was with the Laboratory of Electromagnetics and Antennas (LEMA), Ecole Polytechnique Fédérale de Lausanne (EPFL), 1015 Lausanne, Switzerland (e-mail: mohsen.yousefbeiki@alumni.epfl.ch).}%
\thanks{J. R. Mosig is with the Laboratory of Electromagnetics and Antennas (LEMA), Ecole Polytechnique Fédérale de Lausanne (EPFL), 1015 Lausanne, Switzerland.}%
\thanks{A. Burg is with the Telecommunications Circuits Laboratory (TCL), Ecole Polytechnique Fédérale de Lausanne (EPFL), 1015 Lausanne, Switzerland.}}

\maketitle

\begin{abstract}
Beam-space MIMO has recently been proposed as a promising solution to enable transmitting multiple data streams using a single RF chain and a single pattern-reconfigurable antenna. Since in a beam-space MIMO system radiation pattern of the transmit antenna is exploited as extra dimension for encoding information, near-field interaction of the transmit antenna with its surrounding objects affects spatial multiplexing performance of the system. Through numerical simulations in the previous work, it has been concluded that under BPSK signaling beam-space MIMO is not more vulnerable to near-field coupling than its conventional counterpart. In this work, we extend the study to the case of higher-order modulation schemes, where the presence of external perturbation also affects the data constellation points transmitted by a beam-space MIMO antenna. To this aim, the error vector magnitude of the transmitted signal is evaluated when placing a QPSK beam-space MIMO antenna in close proximity to a hand model of the user. The obtained results emphasize the importance of reconsidering the decoding approach for beam-space MIMO systems in practical applications.
\end{abstract}


%

\section{Introduction}
%
%
%
%
\IEEEPARstart{S}{patial} multiplexing with a single radio is of great interest in wireless communications as an important tool to achieve spectral efficiency in low-end user terminals with stringent complexity and size constraints. To this end, the authors in \cite{kalis2008} introduced the beam-space multiple-input multiple-output (MIMO) concept with the idea of using radiation pattern of the antenna as a knob to \textit{aerially} encode information. More precisely, in beam-space MIMO the instantaneous radiated field of a single-feed reconfigurable antenna is engineered typically by switching variable loads embedded inside the antenna in such a way that multiple symbols are independently mapped onto a predefined orthonormal set of virtual basis patterns in the far-field \cite{alrabadi_tap,my_tap14,my_tcomm}. This idea allows data multiplexing using a single RF chain while remaining compatible with conventional MIMO receivers by maintaining the linear MIMO system model.

Unfortunately, it is well known that the near-field interaction with surrounding objects changes the antenna characteristics such as input impedance and radiation properties. Since a beam-space MIMO system embeds information in the radiation pattern of its transmit antenna, it is expected to be inherently prone to multiplexing performance degradation in the presence of external field perturbation. Hence, the study of the near-field interaction influence on beam-space MIMO is distinct from studies carried out for the case of conventional MIMO systems. 
This issue was first considered in \cite{my_awpl13}, where beam-space MIMO operation under BPSK signaling in the presence of the user body was studied. It has been shown that although the near-field interaction with transmit antennas in beam-space MIMO deteriorates the orthonormality of the basis patterns, throughput reduction is still mainly caused by the power absorption in user body tissues rather than by the distortion of the basis characteristics. While the conclusion in \cite{my_awpl13} confirms the suitability of using beam-space MIMO in near-field interaction scenarios under BPSK modulation, it cannot immediately be generalized to higher-order modulation schemes, where the external perturbation also causes the actual constellation points to deviate from their ideal locations, assumed when decoding under the assumption of a conventional MIMO system model. Such degraded modulation quality can then result in a bit error rate (BER) performance degradation. 

In this letter, beam-space MIMO operation under QPSK signaling in the presence of the user body is analyzed, where the main focus is on the constellation error in beam-space MIMO transmission caused by near-field perturbation and its subsequent impact on the MIMO system performance.

\section{Perturbation Effects in Beam-Space MIMO}
\subsection{Beam-Space MIMO Setup}

\begin{figure}[!t]
\centering
\includegraphics[width=3.9in]{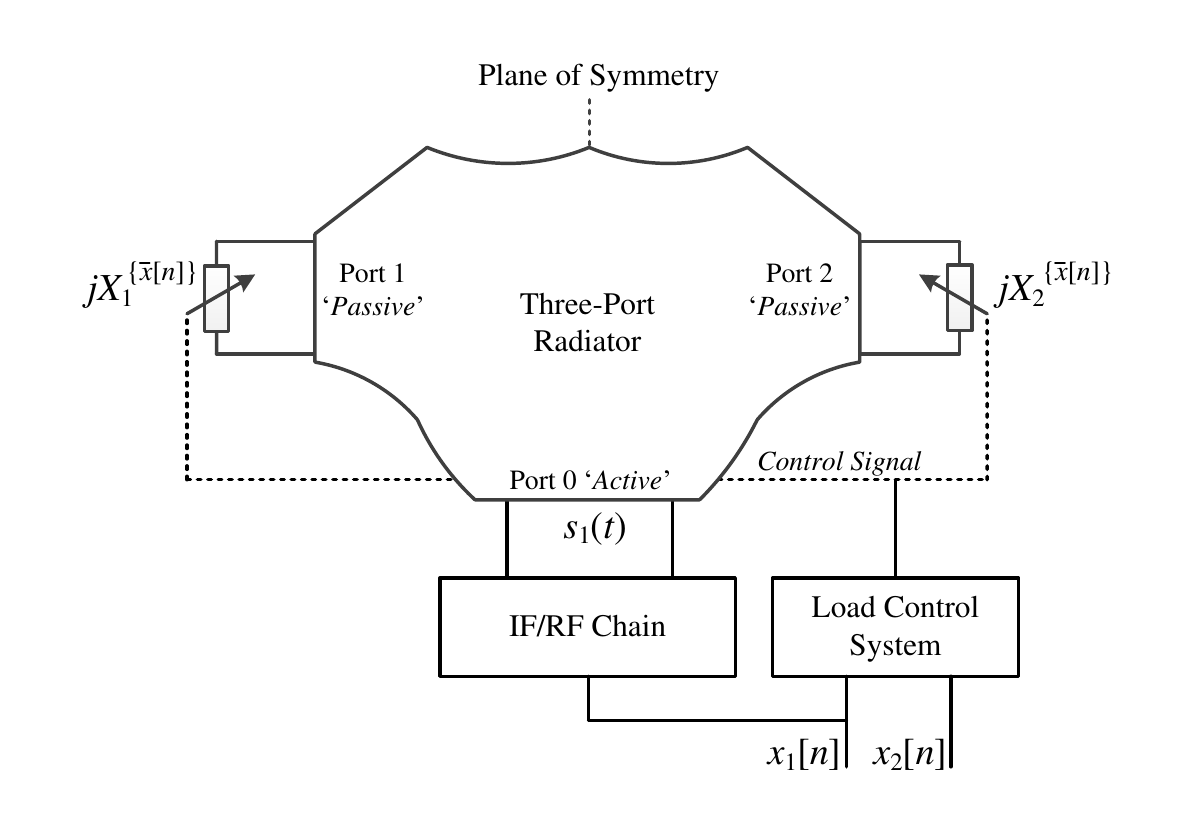}
\caption{Symbolic representation of the single-radio system capable of multiplexing two input symbol streams. $X_p^{\{\bar{x}\}}$ denotes the required reactance value of the variable load at passive port $p$ for a given $\bar{x}$.}
\label{fig:fig1}
\end{figure}

This study is based on the beam-space MIMO antenna introduced in \cite{my_tcomm} and schematically shown in Fig. \ref{fig:fig1}, which allows multiplexing two PSK data streams of any modulation order using a single-feed load-modulated antenna. The antenna is composed of a symmetric three-port radiator and two variable reactive loads, $jX_1$ and $jX_2$, connected to the passive ports of the radiator. The inputs to the system consist of two streams of symbols in the baseband domain, $x_1$ and $x_2$. The first stream, $x_1$, is up-converted to $s_1(t)$, and then fed into the central active port of the radiator. A load control system provides the control signals for reconfiguring the variable loads at the passive ports according to the ratio of two symbols in the baseband domain, namely $\bar{x}= {{{x_2}} \mathord{\left/
 {\vphantom {{{n_2}} {{x_1}}}} \right.
 \kern-\nulldelimiterspace} {{x_1}}}$, such that at each symbol period the actual radiation pattern of the antenna becomes
\begin{align}\label{eq:eq1}
\boldsymbol{\mathcal{E}}_{\circ}^{\{x_1,x_2\}}(\Omega) &= x_1\boldsymbol{\mathcal{E}}_{\rm{u}}^{\{\bar{x}\}}(\Omega)=x_1 \left[ \boldsymbol{\mathcal{B}}_{1}(\Omega) + \bar{x} \boldsymbol{\mathcal{B}}_{2}(\Omega) \right] \\ \nonumber
&= x_1 \boldsymbol{\mathcal{B}}_{1}(\Omega) + x_2 \boldsymbol{\mathcal{B}}_{2}(\Omega) \; ,
\end{align}
where $\Omega$ denotes the solid angle, $\boldsymbol{\mathcal{E}}_{\rm{u}}^{\{\bar{x}\}}(\Omega)$ is the embedded radiation pattern of the antenna (for a unit power excitation), and $\boldsymbol{\mathcal{B}}_{1}(\Omega)$ and $\boldsymbol{\mathcal{B}}_{2}(\Omega)$ represent the basis patterns. In this setup, each symbol stream is independently mapped onto
the corresponding basis pattern in the far-field. Note that all the radiation and basis patterns are $2\times 1$ complex column vectors, where the first and second elements represent the $\hat{\theta}$ and $\hat{\varphi}$ components of the corresponding pattern, respectively.

According to the beam-space MIMO objective given in (\ref{eq:eq1}), the basis patterns can be expressed as
\begin{subequations}\label{eq:eqB}
\begin{align}
\boldsymbol{\mathcal{B}}_{1}(\Omega) &= \frac{1}{2} \left[\boldsymbol{\mathcal{E}}_{\rm{u}}^{\{+1\}}(\Omega)+\boldsymbol{\mathcal{E}}_{\rm{u}}^{\{-1\}}(\Omega) \right]  \\[1mm]
\boldsymbol{\mathcal{B}}_{2}(\Omega) &= \frac{1}{2} \left[\boldsymbol{\mathcal{E}}_{\rm{u}}^{\{+1\}}(\Omega)-\boldsymbol{\mathcal{E}}_{\rm{u}}^{\{-1\}}(\Omega) \right]  \cdot
\end{align}
\end{subequations}
The geometry of the proposed antenna ensures that $\boldsymbol{\mathcal{E}}_{\rm{u}}^{\{-1\}}(\Omega)$ and $\boldsymbol{\mathcal{E}}_{\rm{u}}^{\{+1\}}(\Omega)$ form a mirror image pattern pair when $jX_1^{\{-1\}}=jX_2^{\{+1\}}$ and $jX_2^{\{-1\}}=jX_1^{\{+1\}}$, and therefore guarantees the orthogonality of the basis patterns \cite{my_tcomm}. 

As discussed in \cite{alrabadi_tap}, the signal transmitted by the beam-space MIMO antenna is received using a two-element array. Thanks to the ideal pattern decomposition given in (\ref{eq:eq1}), the beam-space MIMO system model follows the conventional MIMO system model, i.e.,
\begin{equation}\label{eq:eq2}
\mathbf{y}=\left[ {\begin{array}{*{20}{c}}
\hbar_{1,1} & \hbar_{1,2} \\
\hbar_{1,1} & \hbar_{1,2} 
\end{array}} \right] \mathbf{x} + \mathbf{n} \; ,
\end{equation}
where $\mathbf{y}$ and $\mathbf{n}$ are the receive signal vector and the receive noise vector, respectively.
The entries $x_n$ of the transmit symbol vector $\mathbf{x}$ are chosen independently from a set of scalar constellation points $\mathcal{O}$ and $\hbar_{m,n}$ represents the response of the $m$th receive antenna to the $n$th virtual basis pattern. If this model holds, a linear zero-forcing receiver followed by component-wise quantization to the nearest constellation point can recover the two transmitted streams,
\begin{equation}\label{eq:eq3}
\mathbf{\hat{x}}= \mathbf{x} + \left[ {\begin{array}{*{20}{c}}
\hbar_{1,1} & \hbar_{1,2} \\
\hbar_{1,1} & \hbar_{1,2} 
\end{array}} \right]^{-1} \mathbf{n} \; .
\end{equation}

\subsection{Near-Field Perturbation Effects}
Let us assume that the beam-space MIMO antenna has been designed for operation in free space, and there is no mechanism for adapting the reactance values of the variable loads to the immediate environment of the antenna. 

The effects of near-field perturbation on beam-space MIMO operation can be classified into three main levels. In the first level, beam-space MIMO suffers from the typical effects
encountered in any antenna system, namely, the variations of the input impedance and the radiation properties. More precisely, the perturbed radiation pattern can be expressed as
\begin{equation}\label{eq:eff1}
\boldsymbol{\hat{\mathcal{E}}}_{\rm{u}}^{\{\bar{x}\}}(\Omega)= \boldsymbol{\Psi}^{\{\bar{x}\}}(\Omega)\boldsymbol{\mathcal{E}}_{\rm{u}}^{\{\bar{x}\}}(\Omega)\;,
\end{equation}
where $\boldsymbol{\Psi}^{\{\bar{x}\}}(\Omega)$ represents the angular perturbation factor when multiplexing symbol pairs with a ratio of $\bar{x}$.

The second level includes the effects on the characteristics of the basis patterns. According to the definition in (\ref{eq:eqB}), the updated basis becomes 
\begin{subequations}\label{eq:eff2}
\begin{align}
\boldsymbol{\hat{\mathcal{B}}}_{1}(\Omega) &= \frac{1}{2} \left[\boldsymbol{\hat{\mathcal{E}}}_{\rm{u}}^{\{+1\}}(\Omega)+\boldsymbol{\hat{\mathcal{E}}}_{\rm{u}}^{\{-1\}}(\Omega)\right] \\
&= \frac{1}{2} \left[\boldsymbol{\Psi}^{\{+1\}}(\Omega)\boldsymbol{\mathcal{E}}_{\rm{u}}^{\{+1\}}(\Omega)+\boldsymbol{\Psi}^{\{-1\}}(\Omega)\boldsymbol{\mathcal{E}}_{\rm{u}}^{\{-1\}}(\Omega) \right] \nonumber\\[2mm]
\boldsymbol{\hat{\mathcal{B}}}_{2}(\Omega) &= \frac{1}{2} \left[\boldsymbol{\hat{\mathcal{E}}}_{\rm{u}}^{\{+1\}}(\Omega)-\boldsymbol{\hat{\mathcal{E}}}_{\rm{u}}^{\{-1\}}(\Omega)\right] \\
&= \frac{1}{2} \left[\boldsymbol{\Psi}^{\{+1\}}(\Omega)\boldsymbol{\mathcal{E}}_{\rm{u}}^{\{+1\}}(\Omega)-\boldsymbol{\Psi}^{\{-1\}}(\Omega)\boldsymbol{\mathcal{E}}_{\rm{u}}^{\{-1\}}(\Omega)\right]\;.\nonumber
\end{align}
\end{subequations}
Since the perturbation factors $\boldsymbol{\Psi}^{\{+1\}}(\Omega)$ and $\boldsymbol{\Psi}^{\{-1\}}(\Omega)$ are generally unequal, (i) the power radiated by each of two basis patterns is weighted by a different coefficient, and (ii) the perturbed radiation patterns $\boldsymbol{\hat{\mathcal{E}}}_{\rm{u}}^{\{+1\}}(\Omega)$ and $\boldsymbol{\hat{\mathcal{E}}}_{\rm{u}}^{\{-1\}}(\Omega)$ no longer form a mirror image pattern pair. Therefore, not only the basis power imbalance ratio changes, but also the basis orthogonality is no longer guaranteed.

In the third level, near-field interaction of the beam-space MIMO antenna with its immediate environment affects the constellation points mapped over the basis patterns. According to
(\ref{eq:eff1}) and (\ref{eq:eff2}), it can be shown that in general for $\bar{x} \neq \pm 1$,
\begin{align}\label{eq:eff3_1}
\boldsymbol{\Psi}^{\{\bar{x}\}}(\Omega)\boldsymbol{\mathcal{E}}_{\rm{u}}^{\{\bar{x}\}}(\Omega) \neq & \dfrac{1+\bar{x}}{2} \boldsymbol{\Psi}^{\{+1\}}(\Omega)\boldsymbol{\mathcal{E}}_{\rm{u}}^{\{+1\}}(\Omega) \\ 
& +\dfrac{1-\bar{x}}{2} \boldsymbol{\Psi}^{\{-1\}}(\Omega)\boldsymbol{\mathcal{E}}_{\rm{u}}^{\{-1\}}(\Omega) \cdot \nonumber
\end{align}
Therefore, (\ref{eq:eq1}), and consequently the conventional MIMO system model in (\ref{eq:eq2}) no longer hold:
\begin{subequations}\label{eq:eff3_2}
\begin{align}
x_1 \boldsymbol{\hat{\mathcal{E}}}_{\rm{u}}^{\{\bar{x}\}}(\Omega) = x_1 \boldsymbol{\hat{\mathcal{B}}}_{1}(\Omega) + x_2 \boldsymbol{\hat{\mathcal{B}}}_{2}(\Omega) \; {\rm{for}} \; \bar{x}=\pm1 \\
x_1 \boldsymbol{\hat{\mathcal{E}}}_{\rm{u}}^{\{\bar{x}\}}(\Omega) \neq x_1 \boldsymbol{\hat{\mathcal{B}}}_{1}(\Omega) + x_2 \boldsymbol{\hat{\mathcal{B}}}_{2}(\Omega)  \; {\rm{for}} \; \bar{x} \neq \pm1
\end{align}
\end{subequations}
This means that even though the antenna is reconfigured at each symbol period according to $\bar{x}$, the symbols with $\bar{x} \neq \pm1$ are not correctly mapped onto the perturbed basis patterns, and their corresponding constellation points are displaced from their ideal locations when the receiver assumes the conventional MIMO system model in (\ref{eq:eq2}) with $x_n \in \mathcal{O}$. This modulation
distortion can be quantified by the error vector magnitude (EVM) \cite{EVM1}, which for a perturbed beam-space MIMO antenna is a function of the solid angle $\Omega$, given by
\begin{align}\label{eq:evm}
{\rm{EVM}}(\Omega) \!=\!\! \sqrt { {\dfrac{{\displaystyle\sum\limits_{\mathcal{S} = 1}^N \! {\left\| \boldsymbol{\hat{\mathcal{B}}}_{1}(\Omega) \!+\! \bar{x}_{\mathcal{S}} \boldsymbol{\hat{\mathcal{B}}}_{2}(\Omega) \!-\! \boldsymbol{\hat{\mathcal{E}}}_{\rm{u}}^{\{\bar{x}_{\mathcal{S}}\}}(\Omega) \right\|^2}}} {\displaystyle\sum\limits_{\mathcal{S} = 1}^N {\left\| \boldsymbol{\hat{\mathcal{B}}}_{1}(\Omega) + \bar{x}_{\mathcal{S}} \boldsymbol{\hat{\mathcal{B}}}_{2}(\Omega)  \right\|^2}}} } \;,
\end{align}
where $\{\bar{x}_{\mathcal{S}}\}_{\mathcal{S}=1}^N$ are the $N$ possible symbol combination ratios for the considered I-Q constellation points, and $\|\cdot\|$ denotes the Euclidean norm of a vector. Such phase shift and/or amplitude error in turn may result in incorrect interpretation of the signal at the zero-forcing receiver as (\ref{eq:eq3}) is no longer correct.

\section{Simulation Results and Discussion}
\begin{figure}[!b]
\centering
\subfigure[]{\includegraphics[width=1.9in]{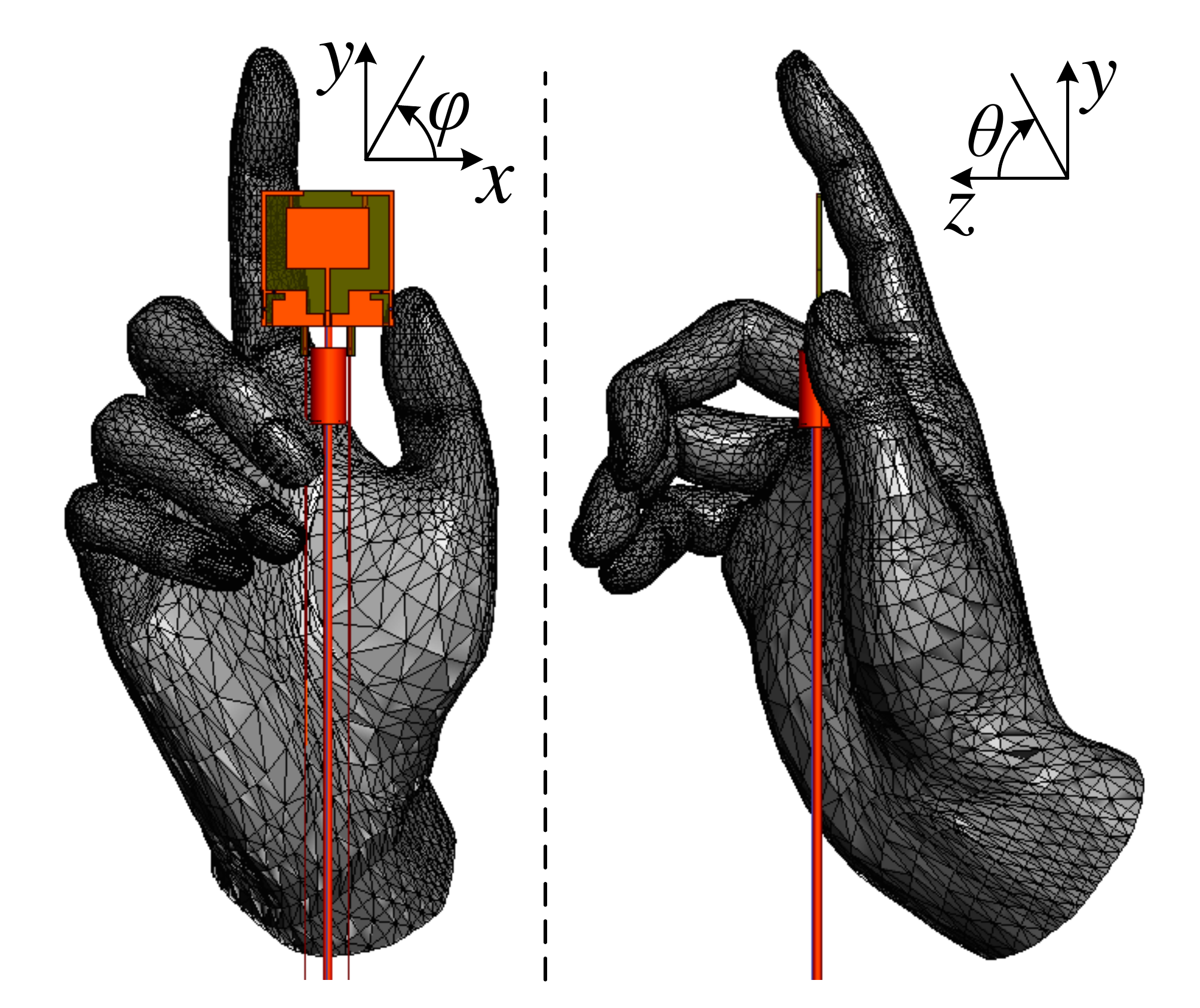} \label{fig:sc1}}
\subfigure[]{\includegraphics[width=2.5in]{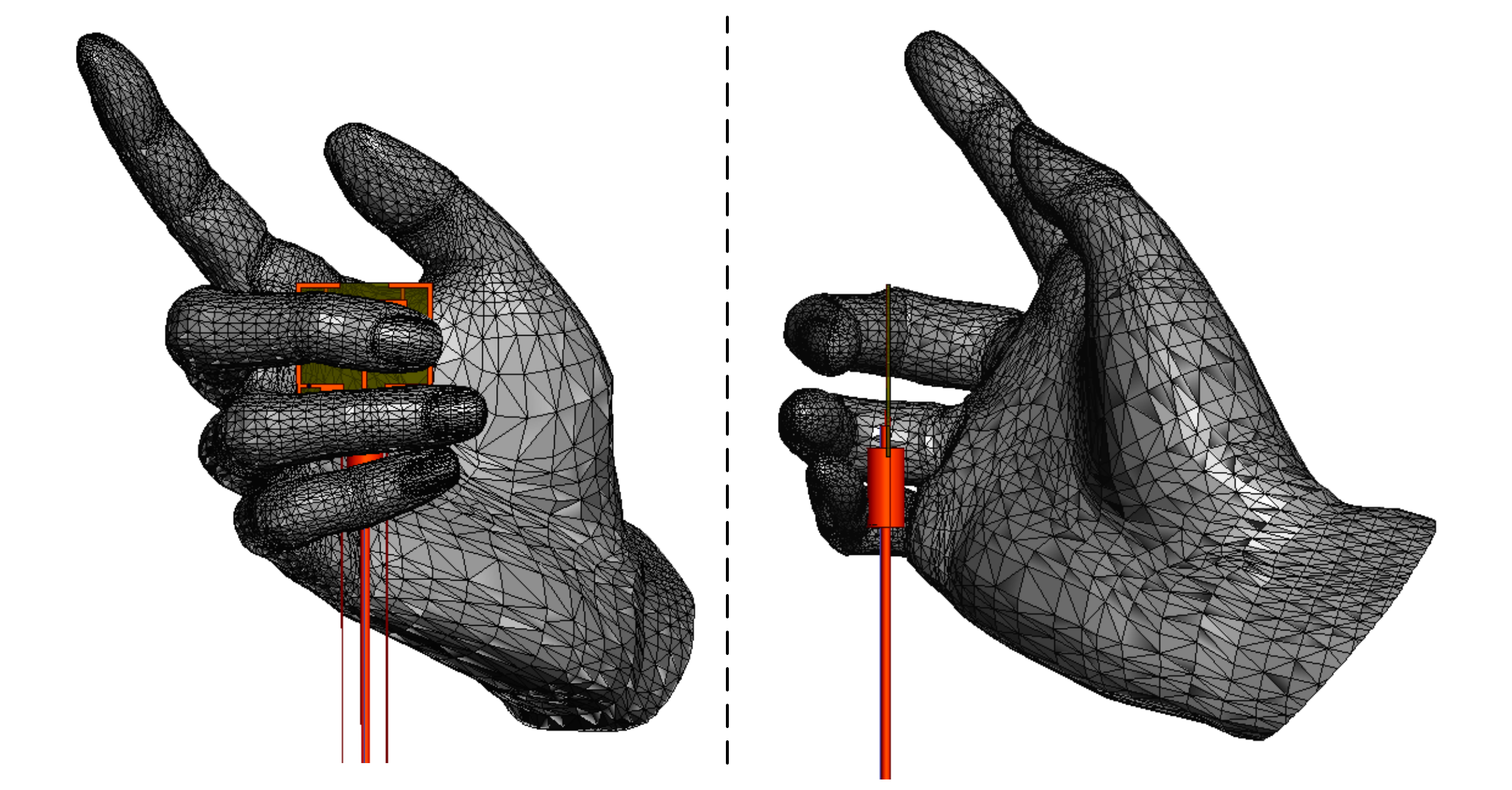} \label{fig:sc2}}
\caption{The placement of the QPSK beam-space MIMO antenna under study near the SAM hand ($\epsilon_r=31.0$, $\sigma=1.64~\rm{S.m^{-1}}$). }
\label{fig:scenarios}
\end{figure}

In this section, we analyze the performance of a QPSK beam-space MIMO antenna in the presence of a full-sized homogeneous specific anthropomorphic mannequin (SAM) hand model in two different scenarios (see Fig. \ref{fig:scenarios}). The full-wave electromagnetic simulations were were carried out using Ansys HFSS. The average permittivity and conductivity of all hand tissues were selected according to the properties of the materials given in \cite{user_permittivity2}. The antenna was originally designed for operation in free space at 2.45 GHz \cite{my_unp1}. Since four distinct values of $\bar{x}$ exist for transmitting two QPSK signals (i.e., $\bar{x}=\{\pm1,\pm j\}$), there are four operational antenna states, and the antenna is capable of creating four radiation patterns according to (\ref{eq:eq1}). 

Table \ref{tab:tab1} summarizes how the characteristics of the beam-space MIMO antenna vary. Obviously, the input impedance and the radiation efficiency are no longer constant over the antenna states in the user body interaction scenarios. The variations of the return loss are rather negligible as the active port of the antenna is still well matched. By contrast, due to the absorption in hand tissues, the radiation efficiency is degraded significantly depending on the antenna state.
Table \ref{tab:tab1} also shows that the power imbalance ratio and the correlation of the basis patterns are affected by the presence of the hand. The power imbalance ratio is degraded from 0.8~dB in free space to 2.1~dB and 5.2~dB in the hand interaction cases. The basis correlation is also increased to $-$11.9~dB in the worst case, which is still sufficiently good for MIMO operation. As concluded in \cite{my_awpl13}, the impairments of the power imbalance ratio and the basis decorrelation show a less pronounced degradation effect on the multiplexing performance of the beam-space MIMO systems.

\begin{table}[t]
\caption{Near-Field Coupling Effect on Beam-Space MIMO Antenna Characteristics}
\vspace{-2mm}
\label{tab:tab1}
\renewcommand{\arraystretch}{0.8}
\centering
\footnotesize
\begin{tabular}{>{\centering}m{2.6cm}>{\centering}m{0.4cm}>{\centering}m{2.2cm}>{\centering}m{2.2cm}>{\centering}m{2.4cm} >{\centering}m{2.2cm}}
\hline \hline \\[-1mm]
Scenario & $\bar{x}$  &  Return Loss (dB) & Radiation Efficiency & Power Imbalance Ratio (dB) & Basis Correlation (dB)\\
\hline \\[-1mm]
 & $-1$ & 19.6 & 75\%  &   & \\[-1mm]
 \multirow{2}{\linewidth}{\centering free space}                           & $+1$ & 19.6 & 75\%  & \multirow{2}{*}{$0.8$} & \multirow{2}{*}{$-\infty$}\\[-1mm]
                            & $+j$ & 19.6 & 75\%  & & \\[-1mm]
                            & $-j$ & 19.6 & 75\%  & & \\[1mm]
\hline  \\[-1mm]
 & $-1$ & 22.5 & 42\%  &   & \\[-1mm]
 \multirow{2}{\linewidth}{\centering in hand, Fig. \ref{fig:sc1}}  & $+1$ & 15.7 & 28\%  & \multirow{2}{*}{$2.1$} & \multirow{2}{*}{$-11.9$}\\[-1mm]
                            & $+j$ & 13.8 & 42\%  & & \\[-1mm]
                            & $-j$ & 24.2 & 27\%  & & \\[1mm]
\hline  \\[-1mm]
 & $-1$ & 12.9 & 26\%  &  & \\[-1mm]
\multirow{2}{\linewidth}{\centering in hand, Fig. \ref{fig:sc2}} & $+1$ & 26.4 & 22\%  & \multirow{2}{*}{$5.2$} & \multirow{2}{*}{$-19.1$} \\[-1mm]
                            & $+j$ & 20.1 & 24\%  &  & \\[-1mm]
                            & $-j$ & 13.3 & 20\%  & & \\[1mm]
\hline    
\end{tabular}
\end{table}

\begin{figure}[!t]
\centering
\includegraphics[width=3.4in]{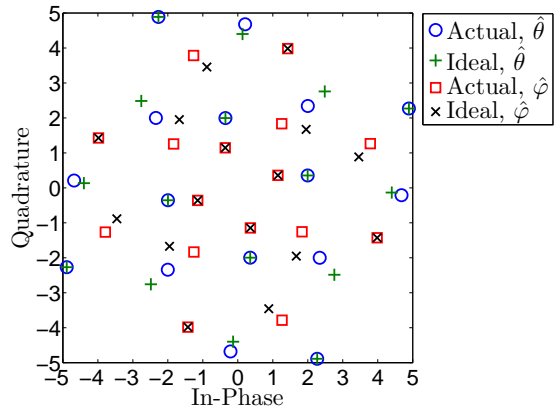} 
\caption{Constellation distortion of the transmitted signal at the solid angle $\{\theta,\varphi\}=\{45^{\circ},294^{\circ}\}$ when placing the antenna near the hand (Fig. \ref{fig:sc1}).}
\label{fig:constellation}
\end{figure}

\begin{figure}[t]
\centering
\subfigure[]{\includegraphics[width=3.9in]{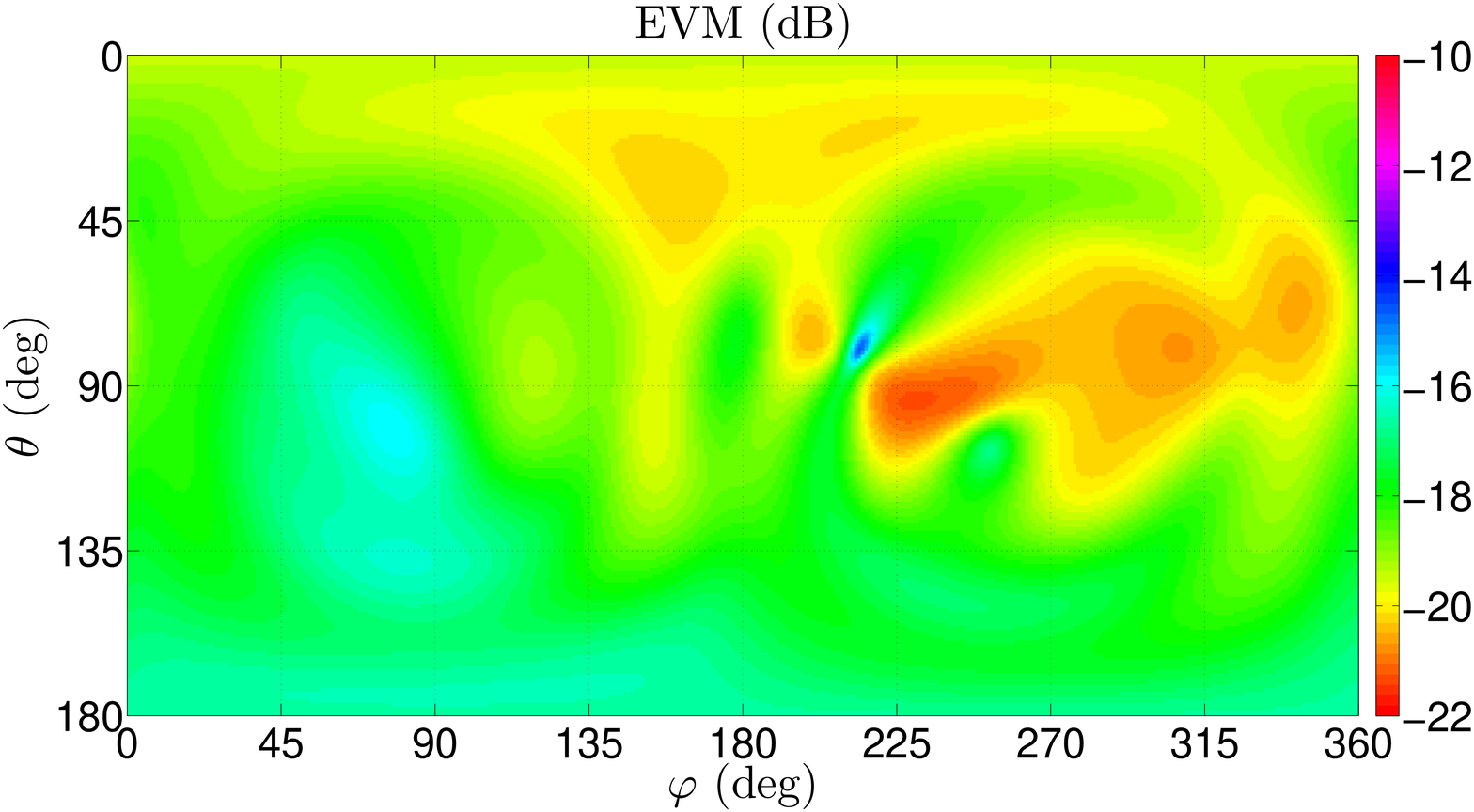} \label{fig:evm1}}
\subfigure[]{\includegraphics[width=3.9in]{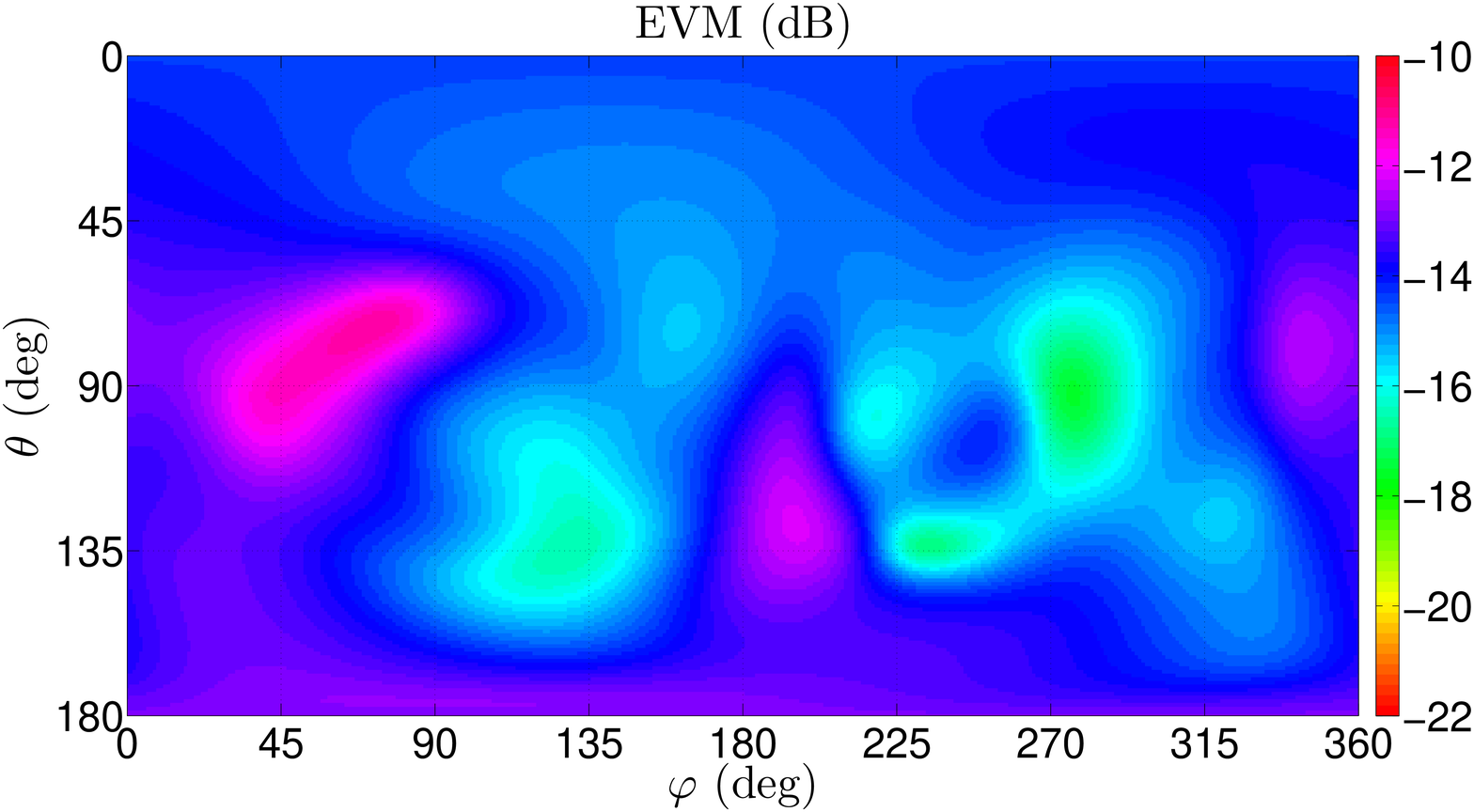} \label{fig:evm2}}
\caption{EVM of the transmitted signal when placing the antenna near the hand; (a) in the case of Fig. \ref{fig:sc1}, and (b) in the case of Fig. \ref{fig:sc2}.}
\label{fig:EVM}
\end{figure}

However, for a receiver that assumes the conventional MIMO input-output relation in (\ref{eq:eq2}) with $x_n \in \mathcal{O}$, it is also important to evaluate the modulation quality of the transmitted signal in both interaction scenarios. Fig. \ref{fig:constellation} shows the constellation diagram of the signal radiated by the perturbed beam-space MIMO antenna at a given solid angle, and compares them with the corresponding points expected from (\ref{eq:eq2}). While the signal has eight of the actual constellation points precisely at the ideal locations, the other eight (corresponding to the symbols with a ratio of $\bar{x}=\pm j$) are displaced. 
Fig. \ref{fig:EVM} shows the EVM of the transmitted signal as a function of the solid angle, calculated using (\ref{eq:evm}). It is observed that the modulation quality is drastically degraded in the entire space. Furthermore, the degradation is highly dependent on the solid angle and the perturbation. Assuming a uniform field distribution (e.g., in a rich-scattering environment), the average EVM is $-$18.5~dB and $-$14.2~dB in the two scenarios, respectively.

\begin{figure}[t]
\centering
\includegraphics[width=3.7in]{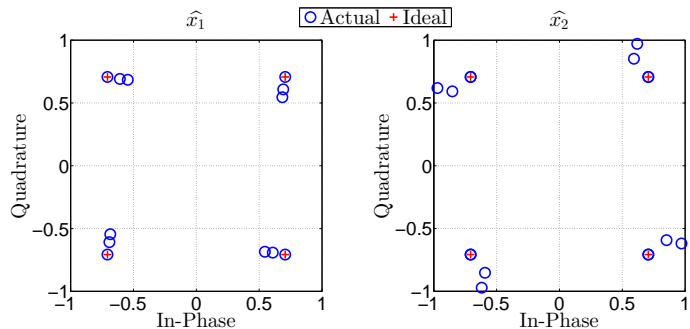} 
\caption{Constellation distortion of the received signals for the receiving antennas at $\{\theta_1,\varphi_1,\theta_2,\varphi_2\}=\{45^{\circ},294^{\circ},45^{\circ},298^{\circ}\}$ when placing the transmit antenna near the hand (Fig. \ref{fig:sc1}).}
\label{fig:cons_RCE}
\end{figure}

\begin{figure}[t]
\centering
\includegraphics[width=3.8in]{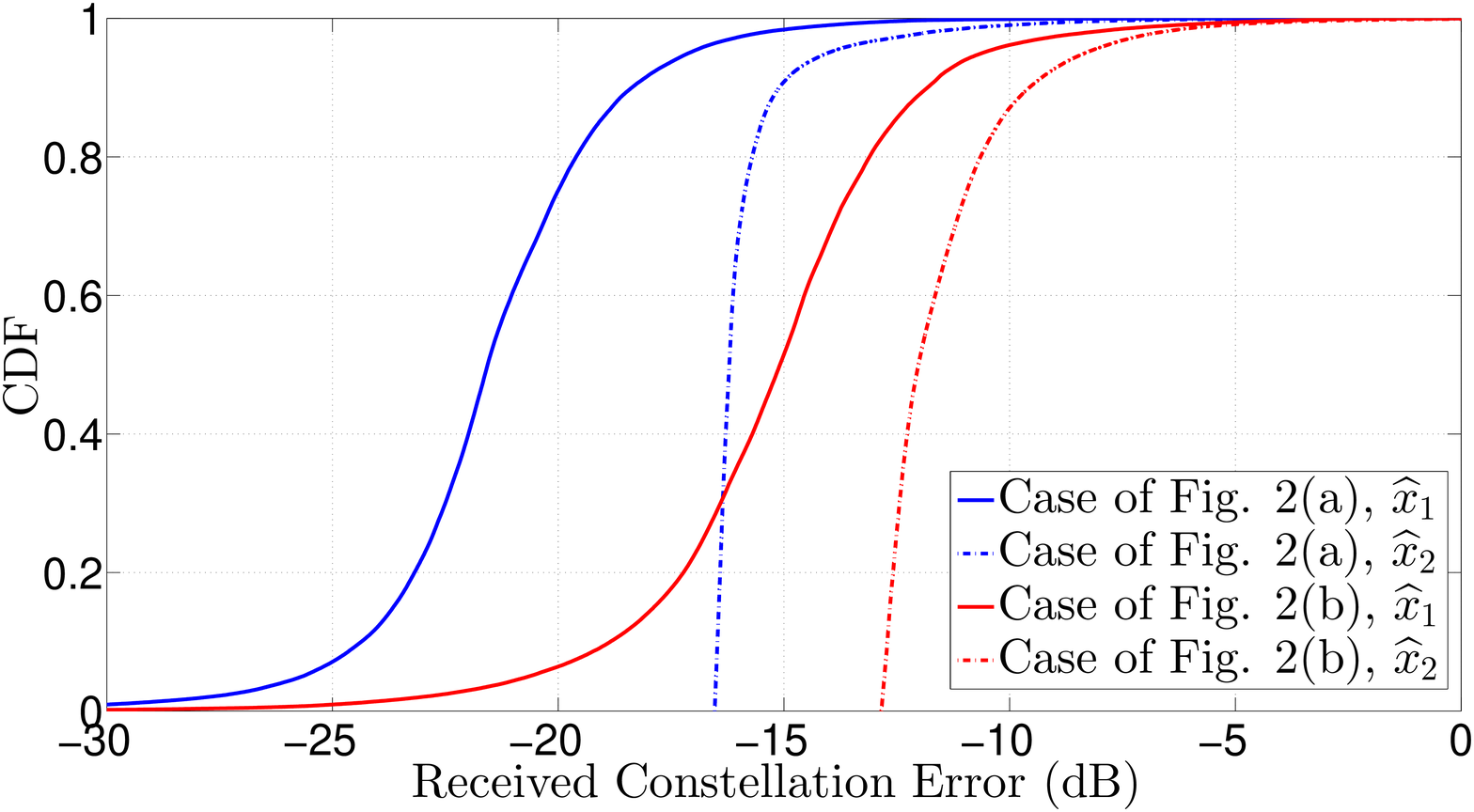} 
\caption{CDF plot of the constellation error of the received signals for 1.4e+6 random single-path LOS transmit-receive scenarios while considering two receive antennas with an angular distance of 3$^{\circ}$-5$^{\circ}$.}
\label{fig:CDF_RCE}
\end{figure}

Finally, to investigate the linear MIMO decoder performance under such EVM degradation, the received constellation error after zero-forcing equalization was calculated for 1.4 million random single-path line-of-sight (LOS) transmit-receive scenarios (while neglecting the noise component). As an example, Fig. \ref{fig:cons_RCE} shows the perturbed constellation diagrams of the equalized symbols for a given transmit-receive scenario. In both diagrams, there are three possible locations for each symbol: the ideal location when the beam-space MIMO antenna radiates in the states corresponding to $\bar{x}=\pm 1$, and the other two when the antenna operates in the states corresponding to $\bar{x}=\pm j$. The cumulative distribution
functions of the constellation error values over different locations in space are depicted in Fig. \ref{fig:CDF_RCE}. It can be seen that the error is more pronounced for the second decoded symbol while depending significantly on the geometry of the near-field interaction scenario. In noisy channels, such errors may result in increased BER of the beam-space MIMO system.


\section{Conclusion}
The paper generalizes a previously reported study for assessing the performance of beam-space MIMO antennas in the presence of external perturbation. It was shown that in the case of higher order modulations, near-field interaction with the beam-space MIMO antenna causes the actual constellation points to deviate from their ideal locations, thereby
increasing the EVM of the transmitted signal. Such degraded modulation quality can then result in incorrect interpretation of the signal with zero-forcing receivers. This emphasizes the importance
of reconsidering the decoding approach for beam-space MIMO systems in user interaction applications.

\section*{Acknowledgment}
The authors would like to dedicate this paper to the memory of Prof. Julien Perruisseau-Carrier.

\ifCLASSOPTIONcaptionsoff
  \newpage
\fi



%




\bibliographystyle{IEEEtran}
\bibliography{IEEEabrv,bibliography}

\begin{thebibliography}{1}
\providecommand{\url}[1]{#1}
\csname url@samestyle\endcsname
\providecommand{\newblock}{\relax}
\providecommand{\bibinfo}[2]{#2}
\providecommand{\BIBentrySTDinterwordspacing}{\spaceskip=0pt\relax}
\providecommand{\BIBentryALTinterwordstretchfactor}{4}
\providecommand{\BIBentryALTinterwordspacing}{\spaceskip=\fontdimen2\font plus
\BIBentryALTinterwordstretchfactor\fontdimen3\font minus
  \fontdimen4\font\relax}
\providecommand{\BIBforeignlanguage}[2]{{%
\expandafter\ifx\csname l@#1\endcsname\relax
\typeout{** WARNING: IEEEtran.bst: No hyphenation pattern has been}%
\typeout{** loaded for the language `#1'. Using the pattern for}%
\typeout{** the default language instead.}%
\else
\language=\csname l@#1\endcsname
\fi
#2}}
\providecommand{\BIBdecl}{\relax}
\BIBdecl

\bibitem{kalis2008}
A.~Kalis, A.~Kanatas, and C.~Papadias, ``A novel approach to {MIMO}
  transmission using a single {RF} front end,'' \emph{IEEE J. Sel. Areas
  Commun.}, vol.~26, no.~6, pp. 972--980, 2008.

\bibitem{alrabadi_tap}
O.~Alrabadi, J.~Perruisseau-Carrier, and A.~Kalis, ``{MIMO} transmission using
  a single {RF} source: Theory and antenna design,'' \emph{{IEEE} Trans.
  Antennas Propag.}, vol.~60, no.~2, pp. 654--664, 2012.

\bibitem{my_tap14}
M.~Yousefbeiki and J.~Perruisseau-Carrier, ``Towards compact and
  frequency-tunable antenna solutions for {MIMO} transmission with a single
  {RF} chain,'' \emph{{IEEE} Trans. Antennas Propag.}, vol.~62, no.~3, pp.
  1065--1073, 2014.

\bibitem{my_tcomm}
M.~Yousefbeiki, O.~Alrabadi, and J.~Perruisseau-Carrier, ``Efficient {MIMO}
  transmission of {PSK} signals with a single-radio reconfigurable antenna,''
  \emph{{IEEE} Trans. Commun.}, vol.~62, no.~2, pp. 567--577, 2014.

\bibitem{my_awpl13}
M.~Yousefbeiki, H.~Najibi, and J.~Perruisseau-Carrier, ``User effects in
  beam-space {MIMO},'' \emph{{IEEE} Antennas Wireless Propag. Lett.}, vol.~12,
  pp. 1716--1719, 2013.

\bibitem{EVM1}
R.~Hassun, M.~Flaherty, R.~Matreci, and M.~Taylor, ``Effective evaluation of
  link quality using error vector magnitude techniques,'' in \emph{Wireless
  Communications Conference}, Aug 1997, pp. 89--94.

\bibitem{user_permittivity2}
C.~Gabriel, ``Tissue equivalent material for hand phantoms,'' \emph{Physics in
  Medicine and Biology}, vol.~52, no.~14, pp. 4205--4210, 2007.

\bibitem{my_unp1}
M.~Yousefbeiki, A.~Austin, J.~R. Mosig, A.~P. Burg, and J.~Perruisseau-Carrier,
  ``Spatial multiplexing of {QPSK} signals with a single radio: antenna design
  and over-the-air experiments,'' submitted to \textit{IEEE Trans. Antennas
  Propag.}.

\end{thebibliography}

%








\end{document}